# Optical Computing with Spectrally Multiplexed Features in Complex Media

Xue Dong[1,*], Kai Lion[2], Fei Xia[1,3], YoonSeok Baek[1], Ziao Wang[1], Niao He[2] and Sylvain Gigan[1, †]

*1. Laboratoire Kastler Brossel, École Normale Supérieure - Paris Sciences et Lettres (PSL) Research University, Sorbonne Université, Centre National de la Recherche Scientifique (CNRS), UMR 8552, Collège de France, 24 rue Lhomond, 75005 Paris, France.*

*2. Department of Computer Science, ETH Zurich, Zurich, Switzerland.*

*3. Department of Electrical Engineering and Computer Science, University of California, Irvine, CA, USA.*

**Abstract**

Artificial intelligence (AI) has rapidly evolved into a critical technology; however, electrical hardware struggles to keep pace with the exponential growth of AI models. Free space optical hardware provides alternative approaches for large-scale optical processing, and in-memory computing, with applications across diverse machine learning tasks. Here, we explore the use of broadband light scattering in free-space optical components, specifically complex media, which generate uncorrelated optical features at each wavelength. By treating individual wavelengths as independent predictors, we demonstrate improved classification accuracy through in-silico majority voting, along with the ability to estimate uncertainty without requiring access to the model's probability outputs. We further demonstrate that linearly combining multiwavelength features, akin to spectral shaping, enables us to tune output features with improved performance on classification tasks, potentially eliminating the need for multiple digital post-processing steps. These findings illustrate the spectral multiplexing or broadband advantage for free-space optical computing.

# Introduction

AI has evolved recently into a critical technology, profoundly impacting various aspects of our daily lives. The innovations in both software algorithms, such as neural networks, and electrical hardware like GPUs have driven the rapid advancement of AI in recent years [1,2]. However, the electronic hardware infrastructure is struggling to keep up with the rapidly increasing computational demands of AI [3]. Modern electronic hardware is largely based on the von Neumann architecture, in which memory and processing units are physically separated. This separation introduces high latency, which limits data transfer rates between memory modules and computational cores and results in inefficiencies during AI training and inference [2]. In addition, the increasing energy consumption of electronic hardware, along with the significant heat generated during processing, presents serious environmental challenges. These issues not only require advanced cooling systems but also raise concerns about sustainability and deployments [4].

To address these issues, researchers have been developing hardware-based AI accelerators to mitigate these bottlenecks on specialized physical systems [5,6], including spintronic [7,8], nano-electronic [9], mechanical [10], photonic [11–14], and free-space optical [15,16] platforms. Among various platforms, free-space optics is particularly effective for in-memory computing, large-scale optical processing, and low-power, low-latency operation [17–20]. In free-space optical computing system, input information is typically encoded using a spatial light modulator (SLM) or digital micromirror device. A physical optical element, such as customized diffusers [17] multimode fibers [21], metasurfaces [22,23] or other complex media [19,24] acts as an optical spatial transformation linear operator, enabling large-scale in-memory optical computation. These architectures take advantage of the inherent massive spatial parallelism of light, providing a natural pathway toward high-throughput processing. Building on these advances, recent research has increasingly explored the use of spectral degrees of freedom to further expand free space optical computational capabilities. Spectral-spatial encoding has recently been applied to increase parallelism and power efficiency [25,26]. High-power broadband optical pulses have been utilized for generating optical nonlinearities for enhancing computational performance [27,28].

Complex media naturally encodes information into both the spatial and spectral domains. This is because light scattering in such a medium depends not only on the spatial position of the input light but also on its wavelength [29]. In linear systems, prior works have typically used monochromatic light and static scattering configurations (e.g., diffusers or multimode fibers) to perform large-scale random projections to bring operations like dimensionality reduction, kernel approximation, and reservoir computing to optical systems [24,30]. In the nonlinear regime, studies have used high-power broadband pulses to induce nonlinear interactions (e.g., second harmonic generation) within the scattering medium, thereby increasing the computational performance [27,28]. Despite these advances, the optical configuration remains fixed in both scenarios, restricting tunability and capability for optical feature extraction.

In this work, we utilized the linear scattering of broadband light in complex media for machine learning applications. Due to the dispersive nature of such media, different wavelengths generate distinct speckle patterns, effectively enabling a set of independent random projections across the spectral domain [29]. This spectral diversity provides an additional degree of freedom and enhances the computational dimensionality of the system. To exploit this, we propose two strategies that utilize the broadband nature of light. The first approach is a spectral ensemble learning method inspired by ensemble learning in machine learning [31–33], where each wavelength provides an independent random projection, and classification is performed by majority voting across their individual predictions. We show that this strategy improves performance and naturally supports uncertainty estimation without requiring access to a model's probability scores. The second strategy leverages the tunability offered by spectral multiplexing, optimizing the final output by combining weighted optical features from different wavelengths, akin to optical spectral shaping. We demonstrate this second approach can maximize task-specific performance. Each method offers distinct advantages in terms of accuracy and efficiency. The added degree of freedom provided by the spectral dimension enhances the flexibility and capacity of optical computing systems, while also being straightforward to integrate into diverse optical hardware architectures. Together, these advantages of optical multiplexing pave the way for more versatile system designs and broader applications of free-space optical computing system.

## Results

### Multiplexing optical features.

Our optical neural network consists of four main components: data encoding, optical feature extraction, intensity detection, and digital readout (Fig 1a, see Fig. S1a for details). Input data is encoded onto the phase of a laser beam using a SLM. The encoded beam then passes through a complex medium (an optical diffuser, see Methods for fabrication details), which scatters the light and generates a speckle pattern. The resulting optical intensity distribution is recorded by a camera.

To further extend the capabilities of this optical system, we introduce wavelength diversity as an additional degree of freedom. The wavelength-tunable laser allows us to illuminate the complex medium at multiple wavelengths. We observe uncorrelated output speckles beyond a 5 nm wavelength difference while keeping the SLM pattern (input data) unchanged (Fig. S1b). This decorrelation arises because the complex medium can be described as a three-dimensional random matrix, incorporating spatial (x, y) and spectral ($\lambda$) variations [29]. The observed drop in correlation between speckle patterns at different wavelengths reflects the medium's spectral response: the interference pattern depends on the path length distribution inside the medium [34]. This wavelength dependent variation enables each wavelength to apply a distinct transformation to the same input data. To utilize this spectral diversity for machine learning, we explore two strategies: spectral ensemble learning, which combines predictions from different wavelengths, and spectral feature optimization, which fuses features across wavelengths into a single optimized representation.

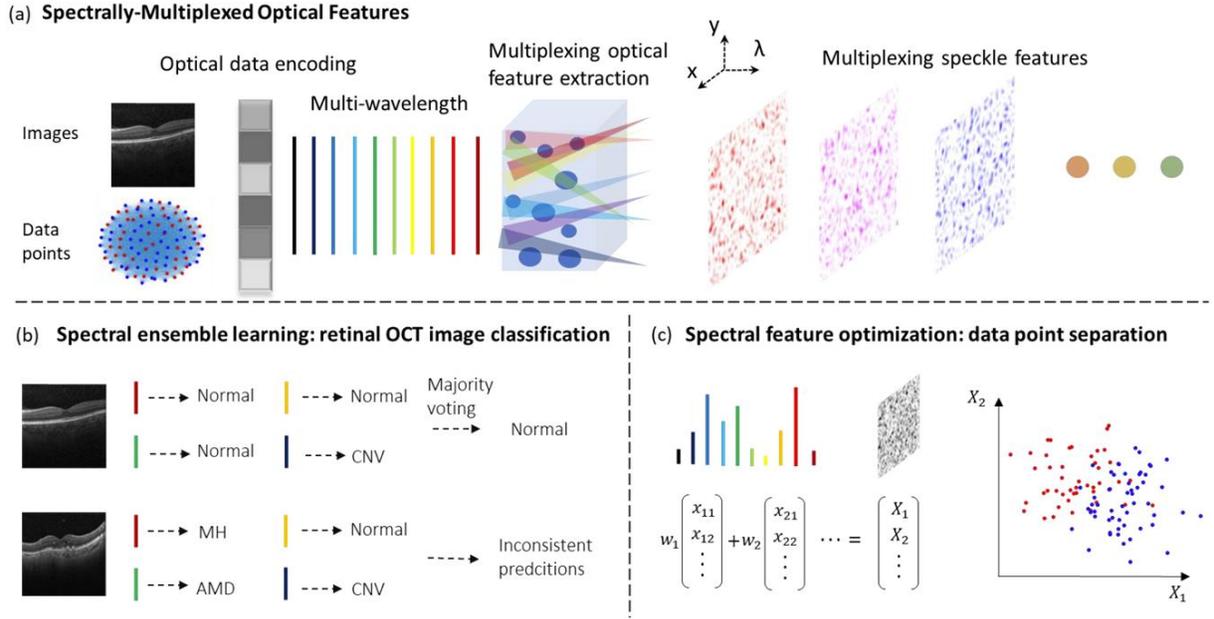

Figure 1. **Multiplexed optical feature extraction with spectral ensemble learning and spectral feature optimization** (a) Schematic setup for feature extraction: The light from a tunable continuous-wave laser is directed onto an optical glass diffuser, producing distinct speckle patterns for each wavelength. (b) Spectral ensemble learning: retinal OCT image classification. Speckles from different wavelengths are processed independently using linear regression. Predictions from each wavelength are combined via majority voting to yield the final classification across eight retinal disease classes. Prediction consistency across wavelengths is used to estimate uncertainty—highly inconsistent results indicate high uncertainty. (c) Spectral feature optimization: data point separation. Speckle patterns from multiple wavelengths are fused using trainable weights to enhance data point separation.

Spectral ensemble learning.

To demonstrate the utility of spectral ensemble learning, we evaluate our optical neural network on a practical task: retinal OCT image classification (Fig. 2a) [35]. As a starting point, we assess the optical neural network's performance at individual wavelengths. This OCT dataset contains 24,000 retinal OCT images categorized into eight retinal disease classes. It consists of 21,200 grayscale training images and 2,800 grayscale testing images, each with dimensions of 224 × 224. Each image is encoded onto the SLM and scattered by the glass diffuser to generate an output feature vector. This process effectively performs a random matrix multiplication in the spatial domain, where each speckle grain corresponds to a unique random feature [5]. By capturing a sufficiently large speckle pattern, a high-dimensional random feature vector is obtained. A digital linear classifier is then trained to predict the final classification results (Fig. 2b) (see Methods). For a regular single-wavelength case, we observe that prediction accuracy increases monotonically with the number of features, consistent with simulation results (Fig. 2f) and prior studies on the use of random features extracted from complex media for approximating kernel functions [24].

Building on these results, we next explore how incorporating multiple wavelengths impacts performance. To assess wavelength-dependent performance, we repeat the above process for wavelengths ranging from 785 nm to 825 nm in 5 nm increments (see Methods), generating 24,000 output samples at each of the nine wavelengths. A separate linear regression model is trained for each wavelength (see Methods), resulting in distinct predictions for the same test image. Prediction accuracy varies slightly across wavelengths, as indicated by the error bars in Fig. 2f. These variations may arise from statistical fluctuations and cannot be taken as strong evidence that one wavelength performs better than another (see Section 1 of the SI). In contrast, the ability to generate distinct predictions across wavelengths opens the door to applying ensemble learning techniques.

Ensemble learning is a widely adopted strategy in machine learning for improving classification accuracy by combining predictions from multiple models [36]. A common approach of the ensemble method is majority voting, where multiple models independently make predictions on the same sample, and the final output is the class predicted by the majority. For majority voting to be effective, two conditions must be met: (1) individual models should perform better than random guessing, and (2) prediction errors from different models are only weakly correlated, meaning they tend to make mistakes on different data points [33].

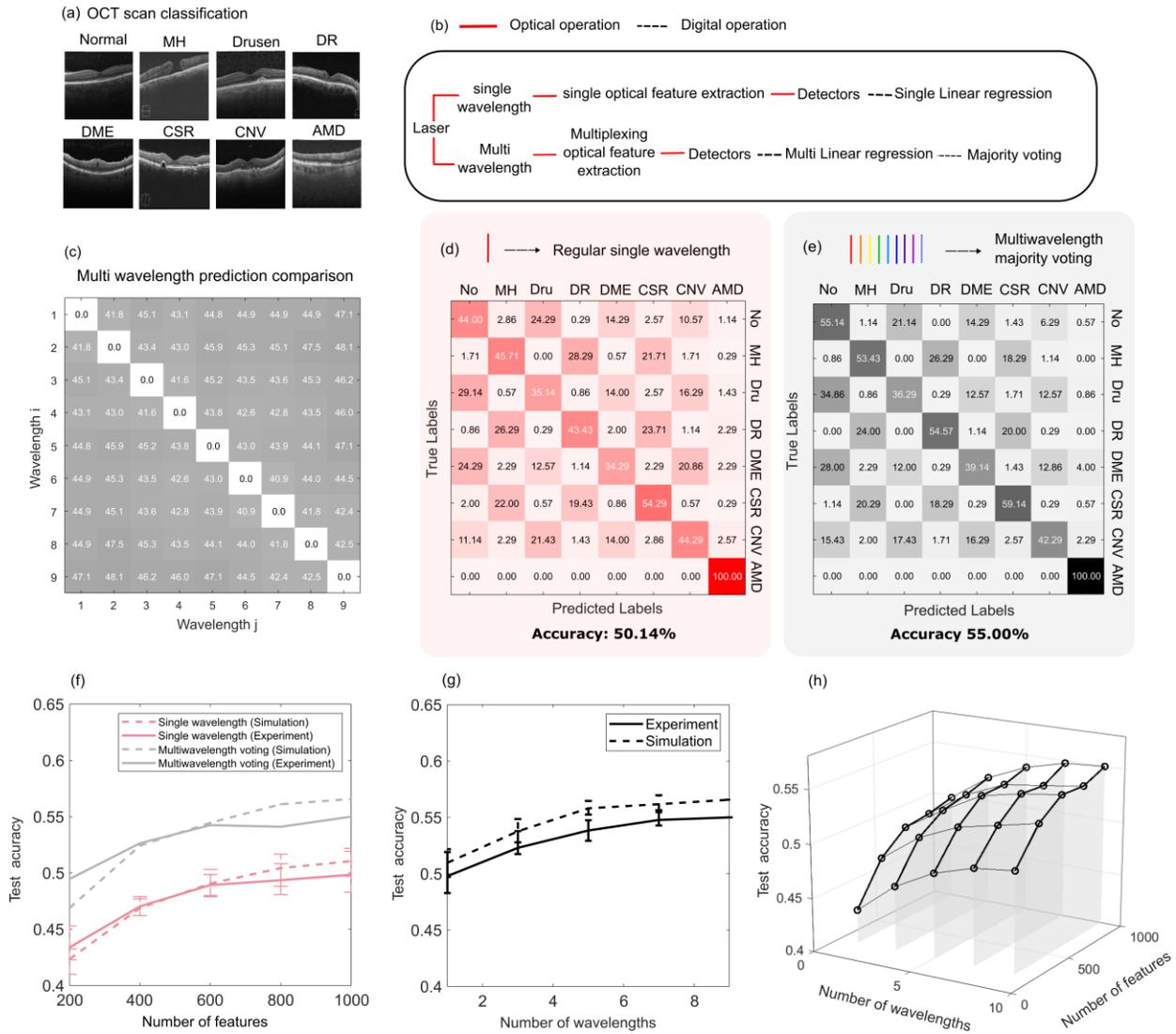

Figure 2. **Experiment of spectral ensemble learning on OCT classification task**: (a) Representative OCT classification images. (b) Logic flow of data collection and processing. (c) Percentage of samples for which predictions from wavelength i differ from those of wavelength j (d) Confusion matrix on the test set using output speckles from a single wavelength at a feature size of 1000. (e) Confusion matrix after combining predictions from all nine wavelengths using majority voting. (f) Comparison of classification accuracies across varying numbers of features: The solid red line shows experimental results from a single wavelength, while the dashed red line shows the corresponding simulations. Prediction accuracy varies slightly across wavelengths, as indicated by the error bars. The solid gray line shows experimental predictions from nine wavelengths combined by majority voting, and the dashed gray

line shows the corresponding simulated results. (g) Classification accuracy at a fixed feature size of 1000 as the number of combined wavelengths increases. Error bars indicate variation across different wavelength groupings. (h) Three-dimensional trend of test accuracy as a function of both number of features and the number of combined wavelengths.

To evaluate these conditions in our system, we analyze the predicting accuracy of single wavelength (Fig. 2d) and pairwise prediction correlations across different wavelengths (Fig. 2c). As shown in Fig. 2d, the single-wavelength model performs better than random guessing. The error bars in Fig. 2f illustrate classification accuracies across different wavelengths, showing that performance consistently exceeds the random baseline across all cases. For reference, random guessing with eight classes would yield an accuracy of only 1/8 (12.5%). To examine the second condition, we calculate the pairwise predictive disagreement between wavelengths. Each cell (i, j) in Fig. 2c shows the percentage of samples for which predictions from wavelength i differ from those of wavelength j. The diagonal entries are zero by definition (indicating perfect self-agreement), while the off-diagonal values were observed ranging from 40% to 48%, indicating substantial variability in model predictions [32]. These findings confirm that both conditions for effective ensemble learning are met, supporting the use of multi-wavelength optical outputs for ensemble-based prediction.

In our setup, each wavelength acts as an independent "voter." We perform linear regression and obtain a prediction result for each wavelength. Each wavelength independently generates predictions on a given test sample, and the final classification result is determined by majority voting expressed as:

$$\hat{y} = \arg\max_{c \in C} \sum_\lambda \mathbb{I}(\hat{y}_\lambda(x) = c) \tag{1}$$

where $C$ represents the set of all possible classes, $\hat{y}_\lambda(x)$ is the prediction from the wavelength $\lambda$ for sample $x$, $\mathbb{I}(\cdot)$ is the indicator function, and $\hat{y}$ is the final predicted class. In this framework, each wavelength contributes one vote toward its predicted class. The summation counts the total number of votes received by class $c$ across all wavelengths. The class with the highest vote count is selected as the final prediction.

Applying majority voting across nine wavelengths increases the classification accuracy to 55.00% from 50.14% at a feature size of 1000 (Fig. 2e). This improvement was observed across different feature sizes (Fig. 2f). We also simulated the majority voting process by applying a unique random matrix for each wavelength to obtain uncorrelated output feature vectors, then performing voting across the results. The simulation outcomes closely match the experimental results (Fig. 2f). This alignment provides strong evidence that the observed performance gain is due to aggregating complementary information across uncorrelated feature representations.

To gain a better understanding of how classification performance is influenced by both the number of wavelengths and feature sizes, we first evaluated accuracy using subsets of 3, 5, and 7 wavelengths selected from the full set of nine. Since multiple combinations are possible, the error bars indicate the variation in prediction accuracy across different wavelength groupings. A clear trend is observed: using more wavelengths generally results in higher test accuracy, as observed in both simulations and experiments (Fig. 2g). Since both larger feature sizes and a greater number of wavelengths contribute to improved performance, we constructed a 3D plot to visualize the combined effects, averaging the results across combinations to remove error bars shown in Fig 2f and Fig.2g (Fig. 2h). The highest accuracy is observed at the largest feature size and all nine wavelengths. Further improvements may be achieved by increasing the number of features or the number of wavelengths (see Discussion).

## Spectral ensemble uncertainty estimation.

The ability to obtain individual predictions from each wavelength not only enhances overall accuracy but also enables a straightforward estimation of predictive uncertainty on the test set without access to the model's probability output. In real-world applications, understanding a model's uncertainty is essential. If the model is uncertain, its predictions should be treated with caution and may need to be referred to alternative models or human experts. In contrast, predictions made with higher confidence are typically more accurate and can be trusted more in practice.

Evaluating predictive uncertainty is not trivial, as there is no ground truth for uncertainty itself. Although probability-based methods are widely used to estimate model uncertainty, implementing such approaches in physical hardware-based AI accelerators can be difficult. In hardware-based AI accelerators, models typically integrated into these systems produce point estimates or class labels, rather than probabilistic outputs [5,19]. Enabling probabilistic outputs often requires architectural modifications or costly additional digital processing steps.

Here, we demonstrate that by introducing wavelength as an additional degree of freedom, our multiplexed optical system naturally produces diverse predictions across different wavelengths without requiring access to internal model probabilities. These per-wavelength predictions enable a straightforward estimation of predictive uncertainty. Specifically, for a given input $x$, we derive an empirical class distribution $\hat{p}(c|x)$, by computing the relative frequency of each predicted class across $M$ wavelengths:

$$\hat{p}(c|x) = \frac{1}{M}\sum_\lambda \mathbb{I}(\hat{y}_\lambda(x) = c) \tag{2}$$

From this empirical distribution, we compute the normalized vote entropy (NVE) [37], which is expressed as (see Methods for derivation):

$$NVE = -\frac{1}{\log(\min(M,|C|))}\sum_{c \in C} \hat{p}(c|x) \log \hat{p}(c|x) \tag{3}$$

Note, when the predictions are consistent across wavelengths, the NVE falls to zero (high confidence); when they are inconsistent, the NVE approaches one (low confidence)

We visualize the distribution of vote entropy (uncertainty) using a histogram and analyze the prediction accuracy within different uncertainty bins (Fig. 3c). The results reveal a strong correlation: samples with lower vote entropy (i.e., higher confidence) are significantly more likely to be correctly classified (see SI, Section 3 for class dependent analysis). This demonstrates that for previously unseen test samples, we can infer the reliability of a prediction based on its estimated uncertainty, even without calibrated probability scores. We further leverage this insight by applying uncertainty thresholding to determine when a prediction should be accepted or flagged for further review (Fig. 3c and 3d). Applying this threshold, we achieved 89.61% accuracy on the subset of test samples that exhibit high confidence (Fig. 3e) (see SI, Section 3 for class dependent analysis).

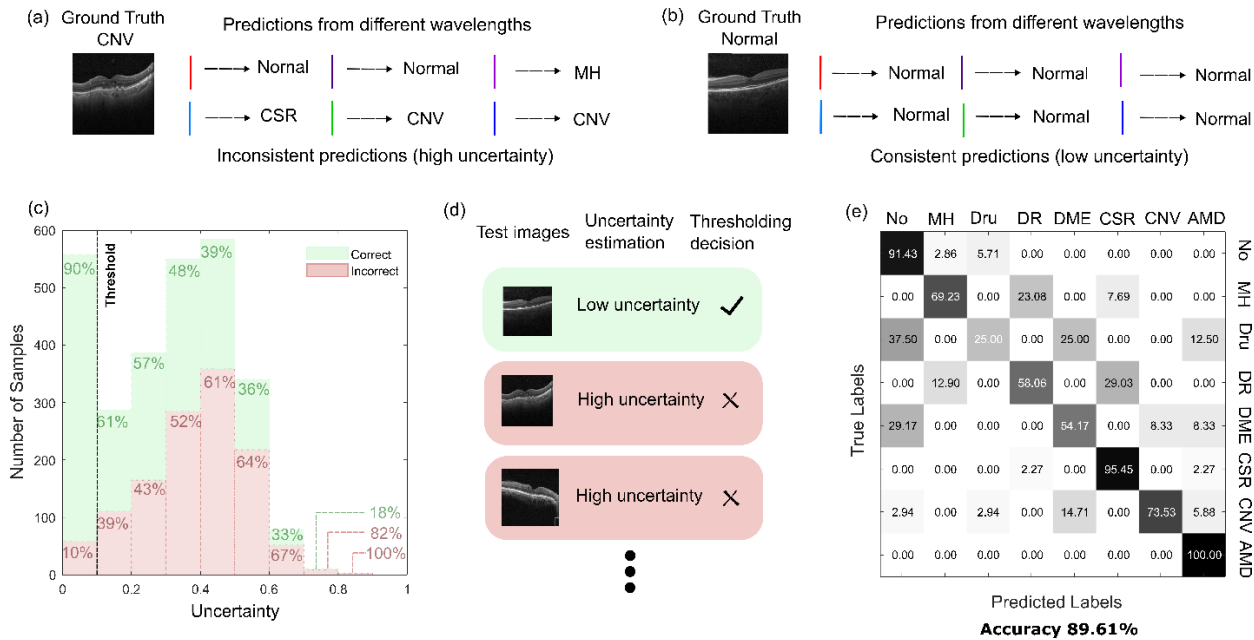

Figure 3. **Spectral ensemble uncertainty estimation on OCT classification task.** Conceptual illustration of (a) inconsistent predictions and (b) consistent predictions across different wavelengths. Different colors of vertical lines indicate different wavelengths. (c) Number of test samples across different uncertainty levels, along with the corresponding percentage counts of correct and incorrect predictions in each uncertainty bin. A threshold is applied at an uncertainty value of 0.1. (d) Conceptual illustration of uncertainty thresholding. (e) Confusion matrix showing classification results for test samples after uncertainty thresholding and majority voting across nine wavelengths.

Summarizing both the spectral ensemble learning and spectral uncertainty estimation, our multiplexed optical system enables both improved classification performance and reliable uncertainty estimation by leveraging diverse predictions across wavelengths. Majority voting improves baseline accuracy, while entropy-based uncertainty estimation allows us to make informed decisions about prediction reliability. By applying thresholding based on uncertainty, we can have high confidence predictions while flagging uncertain cases for further review. This provides a robust and interpretable framework for integrating confidence assessment into real world optical AI systems.

In addition to the advantages offered by multiplexed predictions for improved accuracy and uncertainty estimation, we also explore how the multi-wavelength framework can be extended beyond decision-level fusion. Rather than aggregating predictions after independent processing, we investigate whether the optical features themselves—extracted across different wavelengths—can be directly combined prior to classification. This feature-level fusion enables task-specific optimization of the final feature representation to improve classification accuracy, while maintaining the fast inference time of a single classifier compared to multi-classifier approaches. Moreover, this strategy offers the potential for optical-domain implementation, which has potential for further enhancing the overall system efficiency.

## Spectral feature optimization.

In this approach, we investigate the performance of directly combining optical features into a single fused feature vector and using a single classifier to generate the final prediction, as opposed to merging the final classification outputs from multiple classifiers. In our system, each wavelength produces a distinct random projection of the input, followed by a quadratic non linearity introduced by intensity detection (i.e., the measured output corresponds to the squared magnitude of the optical field). The resulting speckle patterns can therefore be viewed as non-linear features derived from these random projections. When we combine outputs from multiple wavelengths optically, this is equivalent to summing the corresponding intensity vectors. By adjusting the light intensity at each wavelength, we effectively perform a linear combination of these optical features, providing tunability over the final representation (Fig. 4a and Fig. 4b).

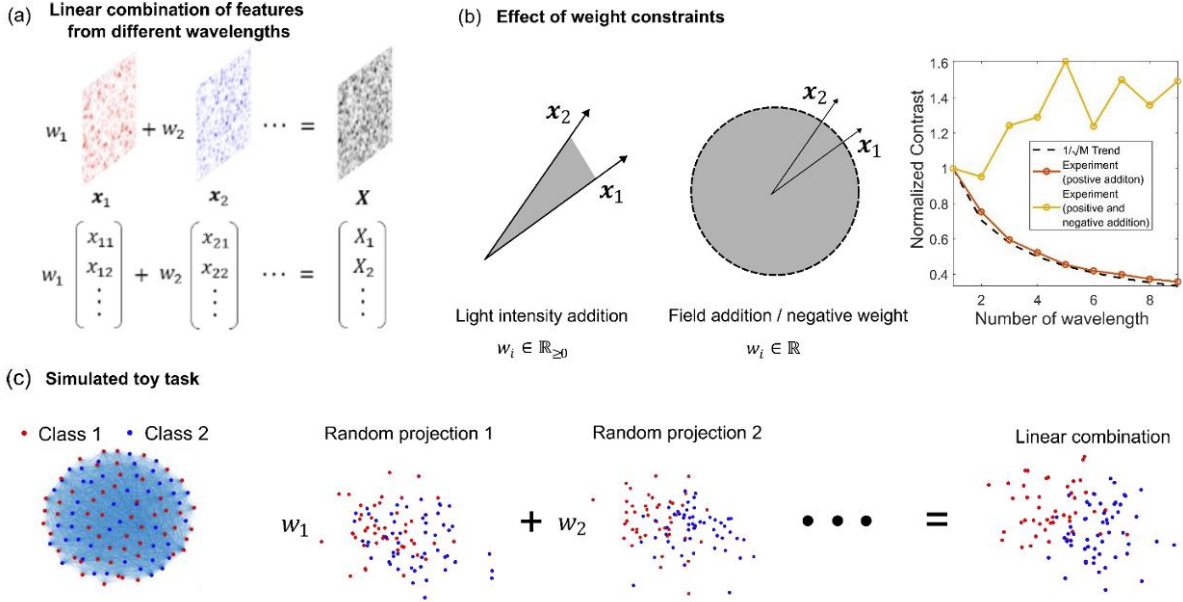

Figure 4. **Spectral feature optimization method** (a) Illustration of the spectral feature optimization method which is a linear addition of the output vectors (b) Comparison of the tunability of linear combinations with and without the non-negativity constraint. Under intensity-based addition, the resulting feature space is limited to a subset of the full 2D span, and the image contrast decreases with the number of wavelengths, following a $1/\sqrt{M}$ trend. In contrast, when both positive and negative weights are allowed, the full linear span of the vectors becomes accessible, and the contrast does not drop with number of wavelengths added. (c) Simulation of toy classification task: Pairwise distances of 100 samples in a 25-dimensional space, divided into two classes. Two examples of a single random projection of the dataset into two dimensions. The projection result after combining nine random projections using a weighted linear combination.

To better understand the nature of this tunable feature space, it is essential to consider the constraints imposed by intensity measurements. Because intensity is inherently non-negative, the resulting output features are strictly positive. If the features were combined optically, it is subject to a non-negativity constraint on the combination coefficients $w_i$ (i.e. $w_i \in \mathbb{R}_{\geq 0}$) (Fig 4a). For example, consider the linear combination of output features at two wavelengths, represented by two uncorrelated vectors $x_1$, and $x_2$. The resulting combination vector $X$ is restricted to a subset of the full 2D plane (Fig. 4b). In a larger space, that will mean the output vector $X$ will lie within the conical hull spanned by all the vectors—also known as a conic combination. In addition to this restriction, summing $M$ number of uncorrelated speckle patterns at the intensity level reduces the speckle contrast by a factor of $1/\sqrt{M}$ (Fig. 4b) [38]. When $M$—in our case, the number of wavelengths—becomes too large, this reduction in contrast can lead to the loss of distinguishable speckle features.

One way to overcome these limitations is to measure the full optical field (amplitude and phase) rather than intensity [38]. Field-based measurement removes the non-negativity constraint and preserves speckle contrast, enabling access to the full feature space and allowing more flexible combinations of optical features. While this approach is optically feasible [39], it typically requires more complex instrumentation and is beyond the scope of the present study. We leave the exploration of field-based implementations for future investigation. In this work, we instead adopt a practical alternative: since speckle patterns at different wavelengths are acquired sequentially, we digitally apply wavelength-dependent weights after acquisition. This allows the use of both positive and negative coefficients, effectively removing the conic constraint and contrast reduction imposed by intensity-only summation (Fig. 4b).

To evaluate the effectiveness of our feature combination approach, we first test it on a synthetic dataset designed for visualization. We generate a two-dimensional, two-class dataset of 100 samples. Class labels are assigned by random sampling, resulting in 46 samples from class 1 and 54 from class 2. Each sample has 25 features, of which only the first two are informative, while the remaining features are noise. Class 1 samples are drawn from a normal distribution centered at [2, 2, 0, …, 0], and class 2 samples are centered at [2, -2, 0, …, 0]; in both cases, noise is added. To visualize the data, we plot the pairwise distances between points (Fig. 4c) (see Methods). We then apply a random projection by multiplying each data point by a random matrix with dimensions matching the data points and a projection dimension of two. Class separation remains generally poor (Fig. 4c). To quantify the separability, we use a linear support vector classifier (SVC), which is well-suited for binary classification tasks. Using the SVC on the training data across nine random projections, we observe a median training accuracy of 70%, with the highest accuracy of 76% (see Methods).

To improve class separability, we perform a linear combination of the random features. To optimize the weights of linear combination, we employ a genetic algorithm that operates by evolving a population of candidate solutions through selection, crossover, and mutation, based solely on the evaluation of a fitness function. As is commonly done in genetic algorithms, we use accuracy as the fitness function. Once the optimal weights are obtained, we plot the final results of the weighted linear combination of the random output features (Fig. 4c). As shown, the classes are better separated, with an accuracy of 91%. These results on this toy task indicate that this method is effective for practical applications where only a few features carry the most relevant information, while others are largely redundant or noisy.

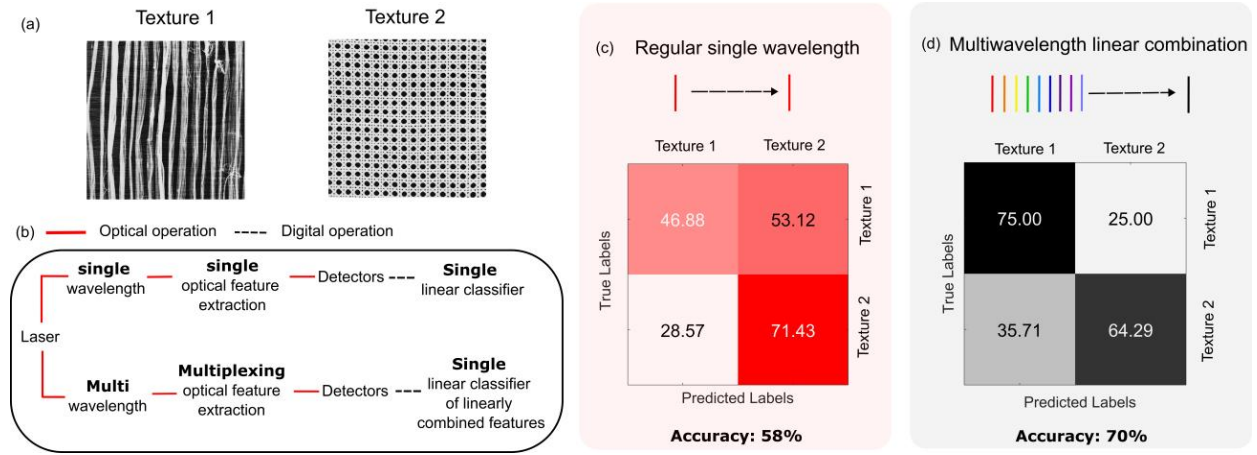

Figure 5. **Experiment of spectral feature optimization on texture classification task** (a) two texture images with 640*640 number of pixels for each image, each image is divided into 15*15 patches. (b) Logic flow of data collection and processing. (c) Confusion matrix of the prediction based on the optical features from a single wavelength and (d) the confusion matrix of the prediction based on linear combined optical features from multiwavelength.

We next apply our method to a real-world classification task. Rather than using the OCT dataset, we evaluate performance on a texture classification task (Fig. 5a), which is more suitable for applying the spectral feature optimization method (see SI section 2). As a starting point, we assess the optical neural network's performance at individual wavelengths. Each texture image has a resolution of 640×640 pixels and is divided into patches (15×15 dimensions, totaling 225 features per patch). We randomly select 150 patches from each texture, resulting in a total of 300 samples (Fig. 5a). These image patches are then displayed on an SLM, and the resulting speckle patterns are recorded. These 300 samples are divided into a training set and a testing set in an 80:20 ratio, resulting in 240 training samples and 60 testing samples. To evaluate single-wavelength performance, we train a separate linear support vector classifier on speckle data from each wavelength independently. At a feature size of 9, we observe a median classification accuracy of 58% across wavelengths (Fig. 5c) (see SI section 2 for the choice of feature size).

Here, we explore how combining multiple wavelengths affects performance. Unlike the synthetic case, which focused solely on training accuracy, we evaluate generalization on a separate testing set here. To determine the optimal linear combination of features across wavelengths, we require a validation set. However, setting aside part of the limited training data for validation would reduce the model's training capacity. To address this, we employ $k$-fold cross-validation method, a common approach in machine learning, to estimate the prediction error using subsets of the training dataset [40]. A genetic algorithm is employed to optimize the weights used in the linear combination across wavelengths (see Methods). After identifying the optimal weights through cross-validation, we apply the learned combination to the held-out test set, which remained completely unseen during both training and validation. We observed an accuracy enhancement reaching 70% from 58% at a feature size of 9 when transitioning from a single wavelength to a linear combination of 9 wavelengths (see SI section 2), indicating the benefit of this method in achieving fast inference due to its single linear classifier

## Discussion

By introducing wavelength as an additional degree of freedom in complex-media-based optical computing systems, we obtain uncorrelated random optical features that can be utilized to enhance learning performance in optical systems. We investigated two strategies for leveraging this spectral diversity: (1) spectral ensemble learning and (2) spectral feature optimization, each offering distinct advantages.

In the spectral ensemble learning approach, predictions from multiple wavelengths are combined at the decision level. This method improves classification accuracy while supporting straightforward uncertainty estimation. Although comparable accuracy gains can be achieved by increasing the number of features collected at a single wavelength, we emphasize that this approach is complementary to spectral multiplexing and the two can be jointly optimized. As shown in Fig. 2h, the highest accuracy is achieved by combining large feature sizes with multiple wavelengths. Accuracy can be further enhanced either by increasing the number of features, which provides a closer approximation to the kernel at a single wavelength [24], or by increasing the number of uncorrelated spectrally multiplexed optical features to have more voters for ensemble. The latter is particularly advantageous for platforms with limited spatial mode, such as multimode fiber–based optical computing systems [21,41]. The number of uncorrelated spectrally multiplexed optical features can be enhanced either with broadband optical sources or with scattering media that have stronger wavelength sensitivity, such as thicker glass diffusers, white paint, or longer multimode fibers. In addition to their role in linear systems, spectrally multiplexed optical features may offer a pathway toward more diverse computational operations: for example, in structural nonlinear cavity systems [42], different wavelengths may experience distinct optical paths or bounce dynamics, enabling richer, wavelength-dependent nonlinear transformations.

In contrast, the spectral feature optimization approach fuses optical features across all wavelengths into a single composite feature vector, which is then processed by a single classifier. This feature-level fusion reduces inference latency by eliminating the need for multiple classifiers, making it well-suited for settings where speed and simplicity are prioritized. However, fusing features optically via intensity addition imposes non-negativity constraints and reduces speckle contrast, limiting the representational capacity. These issues can be mitigated by measuring the optical field (instead of intensity) or applying digital weighting. While this approach offers fewer tunable parameters, it performs particularly well in tasks where the most relevant information is concentrated in a limited number of features and much of the input is redundant. The number of tunable parameters can be increased by increasing the number of uncorrelated spectrally multiplexed optical features.

Overall, introducing wavelength as a degree of freedom opens up new capabilities for optical computing systems. The two strategies we demonstrated address different needs: spectral ensemble learning offers higher accuracy and straightforward uncertainty estimation for complex machine learning tasks at the cost of increased inference time, while spectral feature optimization provides a low-latency solution suitable for simpler tasks. Both approaches are flexible and can be extended to a wide range of optical computing architectures beyond complex media.

## Methods

### Sample information

The optical diffuser used in the experiment was fabricated by sandblasting a coverslip (#1.5) surface with 220-grit aluminum oxide particles. The diffusive surface acts as a wavelength-dependent phase mask due to the geometry of the grooves. Free-space propagation from the diffuser to the camera converts this phase modulation into distinct, wavelength-dependent speckle intensity patterns (see Fig. S1b).

### Experimental setup and data collection

The experimental setup is illustrated in Fig. S1a. A tunable mode-locked Ti: sapphire laser (MaiTai HP, Spectra-Physics) is used as the light source and operates in continuous wave mode. The laser beam passes through a half-wave plate and a polarizing beam splitter to align its polarization with the working axis of the spatial light modulator. A small portion of the beam is directed to a spectrometer (Ocean Optics, HR4000) to monitor the emission wavelength. The main beam is expanded using a 4f system composed of two lenses (L1: f = 12 mm; L2: f = 100 mm), and then directed onto a reflective phase-only SLM (HSP512L-1064, Meadowlark). It has a phase modulation range of [0,π]. The central region of the SLM (approximately 512 × 512 pixels) is used for data encoding. The modulated beam is then transmitted through a glass diffuser and captured by a camera. The camera captures a 150×150 pixel region, which is spatially pooled into a 50×50 speckle map, resulting in a 2500-dimensional feature vector. This process is repeated for wavelengths ranging from 785 nm to 825 nm in 5 nm increments.

Spectral ensemble learning data processing

The multi-wavelength speckle feature vectors are mapped to predicted labels using multiple independent linear regression models. We begin with the optical features obtained from a single wavelength. The output weighted matrix can be analytically expressed as $W_{out} = (K^T K)^{-1} K^T y$ where $K$ is the matrix of speckle feature vectors from the training set of size $N$ and $y$ is the vector of class labels in training set. Because the number of features is much smaller than the number of training samples in our case, no regularization term is required to avoid overfitting. The same training procedure is applied independently at each wavelength. Owing to the wavelength-dependent differences in speckle patterns, each regression produces a distinct $W_{out}$. This generates independent predictions for the same test images across wavelengths. The final classification is obtained by majority voting across all wavelengths, $\hat{y} = \arg\max_{c \in C} \sum_\lambda \mathbb{I}(p_\lambda(x) = c)$. Here, $C$ represents the set of all possible classes, $p_\lambda(x)$ is the prediction from the wavelength $\lambda$ for sample $x$, $\mathbb{I}(\cdot)$ is the indicator function, and $\hat{y}$ is the final predicted class.

     To simulate multiplexing speckle-based computing performance, we generate random complex matrices with Gaussian-distributed entries. The dimensions of each matrix correspond to the input and output feature sizes. To capture the fact that speckle patterns generated at different wavelengths are uncorrelated, we assign a distinct random matrix to each wavelength by initializing the random number generator with a different seed. The simulated speckle features obtained in this way are then processed using the same linear regression training method as in the experiments.

## Entropy calculation

To estimate uncertainty, we start by computing the empirical frequency for each class $\hat{p}(c|x)$ using equation (1). We can then calculate the entropy as a measure of uncertainty.

$$E = -\sum_{c \in C} \hat{p}(c|x) \log \hat{p}(c|x) \quad (4)$$

To ensure that the entropy is scaled between 0 and 1, we normalize it by dividing by the maximum possible entropy. The maximum entropy occurs when the distribution is uniform across all possible classes. If there are $C$ number of distinct classes and $M$ number of available wavelengths, the maximum possible entropy depends on how many unique outcomes we can effectively distinguish in the system. (Even if we have $C$ classes, we cannot distinguish more than $M$ if the data is limited to independent measurements). Therefore, the maximum number of effectively distinguishable outcomes is: $min\ (M, |C|)$ Therefore, we have

$$E_{max} = \log\ (min\ (M, |C|)) \quad (5)$$

So the normalized vote entropy, which serves as our uncertainty estimate can be expressed as equation (3)

## Synthetic data demontration

In Fig. 4c, The node positions in the graph visualization are determined using a force-directed layout, specified by the 'Layout', 'force' option in the plotting function of the MATLAB code. This layout does not use the original coordinates from dataset X, but instead positions nodes based on the structure of the graph built from pairwise Euclidean distances between data points. In this approach, each datapoint becomes a node, and the distance between every pair of nodes is represented as an edge in a fully connected graph. The force-directed layout algorithm models the graph as a physical system where nodes repel each other like charged particles, and edges act like springs that pull connected nodes together. The strength of these springs is influenced by the edge weights, which in this case are the actual distances between datapoints. As a result, datapoints that are closer in the original space tend to be positioned nearer to each other in the graph layout, while those that are farther apart tend to be more separated.

However, it's important to note that this layout only approximates the original distances and does not preserve them exactly. The resulting node positions reflect the relational structure rather than exact geometry.

Unlike real-world datasets, the synthetic dataset used in this toy example is not split into separate training and testing sets. Accordingly, we report training accuracy rather than testing accuracy. The purpose of this toy example is not to assess generalization performance, but to illustrate how the feature combination approach can effectively tune features in a desired direction.

## Contrast calculation

To investigate how the contrast of speckle patterns evolves under different linear combination strategies, we performed two types of speckle summation: purely positive averaging and random signed combination. In the positive addition case, we sequentially combined the first $n$ normalized speckles, where $n$ ranged from 1 to 9. For each value of $n$, the selected speckles were added pixel-wise and divided by $n$ to compute the average speckle image. The contrast of the resulting image was quantified using the standard deviation of its pixel intensities. This procedure models the traditional case of incoherent summation of independent speckle fields. Selected results were visualized to assess qualitative changes in the spatial structure of the speckle patterns as more channels were included.

In the second condition, we studied the behavior of speckle contrast under random signed combinations. For each value of $n$, we performed multiple random trials in which $n$ speckles were randomly selected, and each was assigned a sign of either +1 or −1. The signed speckles were then summed pixel-wise. If the total sum of signs was non-zero, the resulting image was divided by the sign sum to maintain consistency in scaling. If the sum of signs was zero, the raw sum image was used directly to avoid division by zero. The contrast was again measured using the standard deviation. Contrast values were averaged across all trials for each $n$ to obtain a robust estimate.

## Spectral feature optimization data processing

In this approach, we replaced linear regression with a linear SVC, which is well-suited for finding optimal hyperplanes to separate two classes. To integrate information across multiple wavelengths, we employed a linear combination strategy in which feature vectors from all wavelengths were weighted and summed. The optimal set of weights was determined via a genetic algorithm that maximized the average classification accuracy using 5-fold cross-validation on the training set. Once the optimal weights were obtained, they were used to form a combined feature vector for each sample in both training and testing sets. A final linear SVC was trained on the fused features from the training set and evaluated on the combined test set.

## Electrical hardware specification

Throughout all experimental data analysis and simulations, we utilize MATLAB software (MathWorks Inc.) on a computer equipped with an Intel(R) Core(TM) i7-13700 CPU and 32 GB of RAM, one NVIDIA GeForce RTX 3060 GPU, and 28 GB of RAM.

## Data Availability

The raw datasets before optical processing are all publicly available. The recorded experimental speckle feature datasets are available at

https://doi.org/10.5281/zenodo.16955073.

## Code Availability

The code used to produce the results within this work is openly available at

https://doi.org/10.5281/zenodo.16955073.

## Acknowledgements

This work was supported by CRSII5_216600 LION: Large Intelligent Optical Networks. Y.B acknowledges the support from European Union's Horizon Europe research and innovation programme (MSCA-IF N°101105899). S.G. acknowledges support from the Institut Universitaire de France (IUF).

## Corresponding authors

*Corresponding author: xue.dong@lkb.ens.fr

†Corresponding author: sylvain,gigan@lkb.ens.fr

## Competing interests

The authors declare no competing interests.

# Supplementary Information for:
# Optical Computing with Spectrally Multiplexed Features in Complex Media

Xue Dong[1,*], Kai Lion[2], Fei Xia[1,3], YoonSeok Baek[1], Ziao Wang[1], Niao He[2] and Sylvain Gigan[1, †]

1. Laboratoire Kastler Brossel, École Normale Supérieure - Paris Sciences et Lettres (PSL) Research University, Sorbonne Université, Centre National de la Recherche Scientifique (CNRS), UMR 8552, Collège de France, 24 rue Lhomond, 75005 Paris, France.
2. Department of Computer Science, ETH Zurich, Zurich, Switzerland.
3. Department of Electrical Engineering and Computer Science, University of California, Irvine, CA, USA.

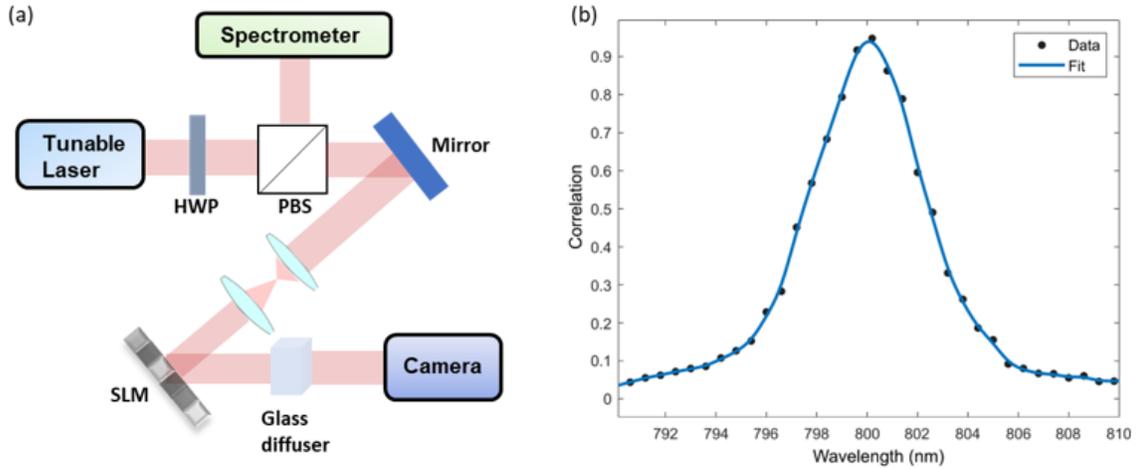

Figure S1. (a) Experimental setup. HWP: half wave plate; PBS: polarizing beam splitter; SLM: spatial light modulator. (b) Spectral optical features correlation of the glass diffuser.

**Section 1: Correlation analysis between training and test accuracy across wavelengths.**

We have demonstrated that different wavelengths produce distinct random projections, resulting in varying classification performance (as shown by the error bars in Fig. 2f). Here, we explore whether this variability can be used to select wavelengths that yield better accuracy. Specifically, we ask: if a particular wavelength achieves higher training accuracy, does it also tend to produce higher accuracy on unseen test data?

To evaluate whether training performance can be used as an indicator for selecting the best performing wavelengths, we compute the cross-validated training accuracy for each wavelength. Cross-validation is a standard technique for estimating how well a model will generalize to unseen data. Instead of training and testing on the same dataset, which risks overestimating performance due to overfitting, the training set is systematically partitioned into multiple subsets, or "folds." In our case, we use 5-fold cross-validation: the data are divided into five equal folds, the classifier is trained on four folds and validated on the remaining one, and this process is repeated so that each fold serves once as validation. The validation accuracies from all folds are then averaged, producing an estimate of how well the model is expected to perform on new samples drawn from the same distribution.

We then compare these cross-validated training accuracies to the actual test accuracies obtained on a completely independent test set that was never used during training. To quantify this relationship, we evaluate the Pearson correlation coefficient (r) and the corresponding p-value. Pearson correlation coefficient measures the strength and direction of a linear relationship between two variables. A value of *r = 1* indicates a perfect positive correlation, *r = −1* a perfect negative correlation, and *r = 0* no correlation. The p-value assesses the statistical significance of this correlation. In our case, we find *r ≈ 0.54*, suggesting a moderate positive trend: wavelengths with higher cross-validated accuracy tend to also yield higher test accuracy. However, $p \approx 0.10$ indicates that there is about a 10% chance that such a correlation could occur randomly. Since this value does not meet the commonly used significance threshold ($p < 0.05$), the evidence is insufficient to conclude that cross-validated accuracy is a reliable predictor of test accuracy for selecting the best-performing wavelength. In other words, it is not sufficient to claim that specific wavelengths possess intrinsic advantages.

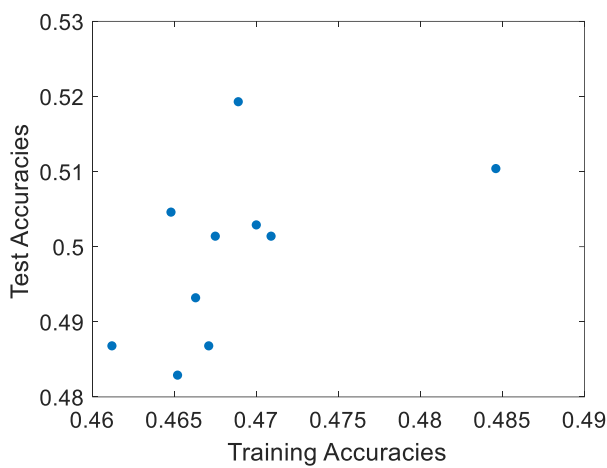

Figure S2. Scatter plot of averaged cross-validated training accuracies versus test accuracies across nine different wavelengths. Each point represents one wavelength condition. A moderate positive trend is visible, with a Pearson correlation coefficient of approximately 0.54. Despite the observed correlation, it is not statistically significant ($p \approx 0.10$).

**Section 2: Task suitability and dimensional constraints of spectral feature optimization.**

We have shown in the main text that spectral feature optimization performs well for tasks characterized by small number of features—such as texture classification. Here, we provide a more detailed explanation of why this approach is inherently better suited for such tasks.

The representational power of spectral feature optimization is fundamentally limited by the number of wavelengths used. Each wavelength provides an independent feature vector $v_i \in \mathbb{R}^D$ and we have M such vectors. We then form the matrix $V \in \mathbb{R}^{M \times D}$ with columns $v_1 \ldots v_M$. For an output vector $y = Vw$, the number of independently controllable components is bounded by $rank(V) = \min(M, D)$. In our system, the number of wavelengths $M$ is much smaller than the feature dimension $D$, so the rank of $V$ is limited to $M$. This constraint implies that the fused feature representation can only explore a low-dimensional subspace of the full feature space. This limitation makes spectral feature optimization less suitable for tasks that require capturing complex, high-dimensional structures. However, for tasks where a small number of features are sufficient to achieve high performance—such as in texture classification—this low-rank representation remains effective. To illustrate this, we set the number of extracted features per wavelength to nine, matching the number of wavelengths used, thereby maximizing control over the final fused representation while maintaining compactness.

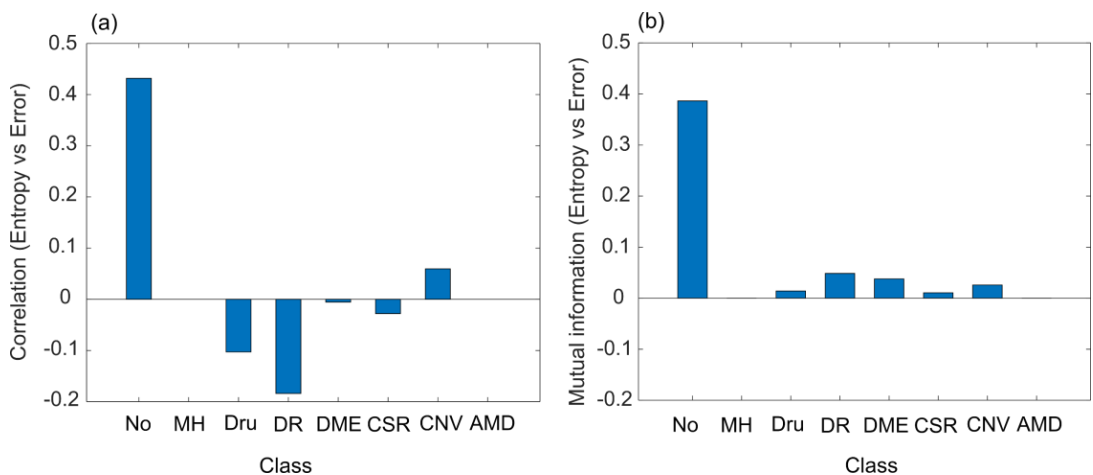

Figure S3. (a) Correlation analysis showing the linear association between uncertainty and error across classes. (b) Mutual information (MI) analysis capturing the full statistical dependence between uncertainty and error.

**Section 3: Analysis of uncertainty–error relationships for each class**

We examined the class dependence of spectral ensemble–based uncertainty estimation by comparing per-class correlations and mutual information (MI) between normalized entropy and prediction wrongness (1 − accuracy). The correlation analysis (Fig. S3a) reveals linear associations between uncertainty and error, whereas the MI analysis (Fig. S3b) captures the full statistical dependence, irrespective of linearity.

Class "normal" exhibited a strong positive correlation (~0.43) and the highest MI (~0.39), indicating that uncertainty is well calibrated in this class: predictions with higher uncertainty are more likely to be incorrect. Several classes showed negative correlations, but MI remained positive, indicating uncertainty still carries predictive information about error, but it is in a nonlinear form not captured by correlation alone.